\documentstyle[12pt]{article}

\def\ket#1{|#1\rangle}

\begin{document}

\title{Shifts on a finite qubit string: A class of quantum baker's maps}

\author{
R\"udiger Schack\thanks{E-mail: r.schack@rhbnc.ac.uk} 
 $^{^{\hbox{\tiny (a,b)}}}$
and
Carlton M. Caves\thanks{E-mail: caves@tangelo.phys.unm.edu} 
 $^{^{\hbox{\tiny (a)}}}$ \\ $\;$ \\
$^{\hbox{\tiny (a)}}$ Center for Advanced Studies, Department of Physics
and Astronomy, \\
University of New Mexico, Albuquerque, NM~87131--1156, USA \\
$^{\hbox{\tiny(b)}}$ Department of Mathematics, Royal Holloway, \\ 
University of London, Egham, Surrey TW20 0EX, UK
}

\date{6 June 1999}
\maketitle

\begin{abstract}
We present two complementary ways in which Saraceno's symmetric version of the
quantum baker's map can be written as a shift map on a string of
quantum bits. One of these representations leads naturally to a
family of quantizations of the baker's map.
\end{abstract}

$\;$ \vspace{1cm}

The main subject of the theory of quantum chaos is the investigation
of {\it quantum signatures of chaos\/} \cite{Haake}, such as characteristic 
eigenvalue statistics \cite{Haake} or hypersensitivity to perturbation 
\cite{hyper}.  In contrast to the situation in classical chaos
theory, many results in the theory of quantum chaos are based on
numerical simulations, not on rigorous proofs. A major part of recent
work on quantum chaos has been the analysis of {\it quantum maps\/}
\cite{Berry}, quantized versions of classically chaotic maps.

Classically chaotic maps are, under very general conditions,
equivalent to Bernoulli shifts on bi-infinite strings of symbols taken
from some finite alphabet. This fact is the basis of the powerful method of
symbolic dynamics \cite{Alekseev1981}, which underlies many of the
rigorous results in classical chaos theory. In the present short paper, we
study shift maps on strings of quantum bits and thus take the first
step towards generalizing the method of symbolic dynamics to the
quantum case.

The quantum baker's map \cite{Balazs1989} is a particularly simple map on the
quantized unit square \cite{Weyl1950}.  It has recently been shown
\cite{Schack1998a,BrunSchack} to have an experimental realization on
present-day quantum computers \cite{Cory1997,Gershenfeld1997}. Here
we investigate finite shift-map representations of the quantum baker's
map.  Shifts on infinite quantum spin chains in the context of
quantum chaos have been discussed in \cite{Alicki1994}. A related
symbolic description of the quantum baker's map is given in
{\cite{Saraceno1994a}}.

The classical baker's transformation \cite{Arnold1968}, which maps the
unit square $0 \leq q,p \leq 1$ onto itself, has a simple description
in terms of its symbolic dynamics \cite{Alekseev1981}. Each point in
phase space is represented by a symbolic string 
\begin{equation}
s = \cdots s_{-2} s_{-1} s_0 . s_1 s_2 \cdots \;,
\label{eqbakerstring}
\end{equation}
where $s_k=0$ or $1$. The string $s$ is identified with a point $(q,p)$
in the unit square by setting 
\begin{equation}
q=\sum_{k=1}^{\infty}s_k 2^{-k}
\label{eqq}
\end{equation}
and
\begin{equation}
p=\sum_{k=0}^{\infty}s_{-k} 2^{-k-1} \;.
\label{eqp}
\end{equation}
The action of the baker's map on a
symbolic string $s$ is given by the shift map $U$ defined by $Us=s'$,
where $s'_k=s_{k+1}$. This 
means that, at each time step, the entire string is shifted one place to
the left while the dot remains fixed.  Geometrically, if $q$
labels the horizontal direction and $p$ labels the vertical, the
baker's map on the unit square is equivalent to stretching the $q$ direction
and squeezing the $p$ direction, each by a factor of two, and then
stacking the right half on top of the left.

We now quantize the unit square
as in \cite{Weyl1950,Saraceno1990}. To represent the unit square
in $D$-dimensional Hilbert space, we start with unitary ``displacement''
operators $\hat U$ and $\hat V$, which produce displacements in the
``momentum'' and ``position'' directions, respectively, and which obey the
commutation relation \cite{Weyl1950}
\begin{equation}
\hat U\hat V = \hat V\hat U\epsilon \;,
\end{equation}
where $\epsilon^D=1$. We choose $\epsilon=e^{2\pi i/D}$.  For consistency 
of units, we let the quantum scale on ``phase space'' be $2\pi\hbar=1/D$.  We
further assume that $D=2^N$, which is the dimension of the Hilbert space of $N$
qubits (i.e., $N$ two-state systems), and that $\hat V^D=\hat U^D=-\hat1$. 
The latter choice enforces antiperiodic boundary conditions; it is motivated 
by the fact {\cite{Saraceno1990}} that for an even dimension $D$, antiperiodic 
boundary conditions guarantee that the classical and quantized maps have 
similar symmetry properties. For an alternative quantization using periodic 
boundary conditions, see \cite{RubinSalwen}. It follows 
\cite{Weyl1950,Saraceno1990} that the operators $\hat U$ and $\hat V$
can be written as
\begin{equation}
\hat U=e^{(i/\hbar)\hat q/D}=e^{2\pi i\hat q}
\qquad\mbox{and}\qquad 
\hat V=e^{-(i/\hbar)\hat p/D}=e^{-2\pi i\hat p} \;,
\end{equation}
where the ``position'' operator $\hat q$ has eigenvalues $q_j=(j+{1\over2})/D$, 
$j=0,\ldots,D-1$, and likewise the ``momentum'' operator $\hat p$ has
eigenvalues $p_k=(k+{1\over2})/D$, $k=0,\ldots,D-1$.

The $D=2^N$ dimensional Hilbert space modeling the unit square can be 
realized as the product space of $N$ qubits in such a way that
\begin{equation}
\ket{q_j} =
\ket{x_1}\otimes\ket{x_2}\otimes\cdots\otimes\ket{x_N}  \;,
\end{equation}
where $j=\sum_{l=1}^N x_l2^{N-l}$, $x_l\in\{0,1\}$,
and where each qubit has basis states $|0\rangle$ and $|1\rangle$.
It follows that, written as binary numbers, $j=x_1x_2\ldots x_N$ and 
$q_j=0.x_1x_2\ldots x_N1$. We define the notation 
\begin{equation}
\ket{.x_1x_2\ldots x_N} = e^{i\pi/2} \ket{q_j} \;,
\label{eqtensor}
\end{equation}
which is closely analogous to 
Eq.~(\ref{eqbakerstring}), where the bits to the right of the dot specify 
the position variable; the reason for the phase shift $e^{-i\pi/2}$ becomes
apparent below.

Momentum and position eigenstates are related through the quantum Fourier
transform operator $\hat F$ {\cite{Saraceno1990}}, defined by
$\hat F\ket{q_k}\equiv\ket{p_k}$, where
\begin{equation}
\ket{p_k} = 
{1\over\sqrt D}\sum_{j=0}^{D-1}\ket{q_j}e^{(i/\hbar)p_kq_j} =
{1\over\sqrt{2^N}}\sum_{x_1,\ldots,x_N}\ket{x_1}\otimes\ldots\otimes\ket{x_N} 
\,e^{2\pi iax/2^N} \;.
\end{equation}
In this expression $a=k+{1\over2}=a_1\ldots a_N.1=2^N p_k$, and
$x=j+{1\over2}=x_1\ldots x_N.1=2^N q_j$. We now define the notation
\begin{equation}
\ket{a_N\ldots a_1.} = \ket{p_k} \;,
\end{equation}
which is again analogous to
Eq.~(\ref{eqbakerstring}), where the bits to the left of the dot, read
backwards, specify the momentum variable.

It will be useful to define a {\it partial\/} Fourier transform, $\hat G_n$, 
which acts on the $N-n$ least significant bits of a state,
\begin{eqnarray}
&\mbox{}& \hat G_n\,\ket{x_1}\otimes\ldots\otimes\ket{x_n}\otimes
\ket{a_1}\otimes\ldots\otimes\ket{a_{N-n}} \nonumber\\
&\mbox{}& \hphantom{G_n|}
=\ket{x_1}\otimes\cdots\otimes\ket{x_n}\otimes
 {1\over\sqrt{2^{N-n}}}\sum_{x_{n+1},\ldots,x_N}\ket{x_{n+1}}\otimes
\cdots\otimes\ket{x_N}
 \,e^{2\pi iax/2^{N-n}}\qquad \;
\label{eqpartial}
\end{eqnarray}
where now $a$ and $x$ are defined by the binary expansions 
$a=a_1\ldots a_{N-n}.1$ and $x=x_{n+1}\ldots x_N.1$.  Again in close
analogy to Eq.~(\ref{eqbakerstring}), we define the notation
\begin{equation}
\ket{a_{N-n}\ldots a_1.x_1\ldots x_n}
= \hat G_n\,\ket{x_1}\otimes\ldots\otimes\ket{x_n}\otimes
\ket{a_1}\otimes\ldots\otimes\ket{a_{N-n}}
\end{equation}
Notice that had we 
instead used $x=x_1\ldots x_N.1$ in Eq.~(\ref{eqpartial}), the only 
difference would have been to multiply $\ket{a_{N-n}\ldots a_1.x_1\ldots x_n}$ 
by a phase $e^{i\pi x_n}$.  The operator $\hat G_n$ is unitary, and the states 
$\ket{a_{N-n}\ldots a_1.x_1\ldots x_n}$ form an orthonormal basis.
As our notation requires, for $n=0$ the states 
$\ket{a_{N-n}\ldots a_1.x_1\ldots x_n}$ reduce to the momentum eigenstates 
(i.e., $\hat G_0=\hat F$), and for $n=N$ they reduce to 
$e^{i\pi/2}\ket{x_1}\ldots\ket{x_N}=|.x_1\ldots x_N\rangle$ 
(i.e., $\hat G_N=i\hat 1$).  The phase shift for $n=N$ is the reason for the
$\pi/2$ phase shift in Eq.~(\ref{eqtensor});
it is a consequence of the antiperiodic boundary conditions.

The state $\ket{a_{N-n}\ldots a_1.x_1\ldots x_n}$ is localized in both 
position and momentum: it is strictly localized within a position region 
of width $1/2^n$, centered at position 
$q=0.x_1\ldots x_n1$, and it is crudely localized within 
a momentum region of width $1/2^{N-n}$, centered at momentum 
$p=0.a_1\ldots a_{N-n}1$.  Using the notation of 
Eq.~(\ref{eqbakerstring}) for phase-space points, we can say that the 
states $\ket{a_{N-n}\ldots a_1.x_1\ldots x_n}$ are localized near the points
$1a_{N-n}\ldots a_1.x_1\ldots x_n1$, with position and momentum widths 
determined by this lattice of points.  

The quantum baker's map as defined in {\cite{Saraceno1990}} is now given by 
{\cite{Schack1998a}}
\begin{equation}
\hat B = \hat G_0 \circ \hat G_1^{-1}\;. 
\label{eqqbaker}
\end{equation}
By noting that 
\begin{equation}
\hat G_0\ket{x_1}\otimes\ket{a_1}\otimes\ldots\otimes\ket{a_{N-1}}=
\ket{a_{N-1}\ldots a_1x_1.}
\end{equation}
and
\begin{equation}
\hat G_1\ket{x_1}\otimes\ket{a_1}\otimes\ldots\otimes\ket{a_{N-1}}=
\ket{a_{N-1}\ldots a_1.x_1}\;,
\end{equation}
one sees that the action of the baker's map is equivalent to shifting the dot
in the symbolic representation, i.e.,
\begin{equation}
\hat B\ket{a_{N-1}\ldots a_1.x_1} = \ket{a_{N-1}\ldots a_1x_1.} \;, 
\end{equation}
similar to the classical symbolic dynamics~(\ref{eqbakerstring}).   Motivated
by this form for the quantum baker's map and by the symbolic representation
of Eq.~(\ref{eqbakerstring}), we can define a whole class of quantum baker's 
maps, $\{\hat B_n\mid n=1,\ldots,N\}$, through
\begin{equation}
\hat B_n\ket{a_{N-n}\ldots a_1.x_1\ldots x_n} 
= \ket{a_{N-n}\ldots a_1x_1.x_2\ldots x_n} \;.
\label{eqgeneralbaker}
\end{equation}
In phase-space language, the map $\hat B_n$ takes a state localized at 
$1a_{N-n}\ldots a_1.x_1\ldots x_n1$ to a state localized at 
$1a_{N-n}\ldots a_1x_1.x_2\ldots x_n1$, while it stretches the state by
a factor of two in the $q$ direction and squeezes it by a factor of two 
in the $p$ direction.

The classical shift map acting on the symbolic string~(\ref{eqbakerstring}) 
can be regarded equivalently as either a right-shift of the dot or a 
left-shift of the infinite string of bits. We now show that, complementary
to the dot-shifting representation (\ref{eqgeneralbaker}), there is a 
representation of the quantum baker's map $\hat B_n$ as a shift of the 
qubits.  Following \cite{Cleve1998}, we write the partial Fourier 
transform~(\ref{eqpartial}) as a product state
\begin{eqnarray}
\label{eqprod1}
&\mbox{}&\hspace{-20pt}
\ket{a_{N-n}\ldots a_1.x_1\ldots x_n} 
\nonumber \\
&\mbox{}&\hspace{-20pt}\hphantom{\ket{a_N}}
=\ket{x_1}\otimes\ket{x_2}\otimes\cdots\otimes\ket{x_n} 
\nonumber \\
&\mbox{}&\hspace{-20pt}\hphantom{\ket{a_{N-n}\ldots}}
\otimes{1\over\sqrt2}\Bigl(\ket{0}+e^{2\pi i(0.a_{N-n}1)}\ket{1}\Bigr)
\otimes 
{1\over\sqrt2}\Bigl(\ket{0}+e^{2\pi i(0.a_{N-n-1}a_{N-n}1)}\ket{1}\Bigr) 
\nonumber \\
&\mbox{}&\hspace{-20pt}\hphantom{\ket{a_{N-n}\ldots}}
\otimes \cdots \otimes
{1\over\sqrt2}\Bigl(\ket{0}+e^{2\pi i(0.a_1\ldots a_{N-n}1)}\ket{1}\Bigr) 
e^{i\pi(0.a_1\ldots a_{N-n}1)}
\;.
\nonumber \\
\end{eqnarray}
Similarly, we can write
\begin{eqnarray}
\label{eqprod2}
&\mbox{}&\hspace{-25pt}
\ket{a_{N-n}\ldots a_1x_1.x_2\ldots x_n} 
\nonumber \\ 
&\mbox{}&\hspace{-25pt}\hphantom{\ket{a_n}}
=\ket{x_2}\otimes\cdots\otimes\ket{x_n} 
\nonumber \\    
&\mbox{}&\hspace{-25pt}\hphantom{\ket{a_{N-n}\ldots}}
\otimes{1\over\sqrt2}\Bigl(\ket{0}+e^{2\pi i(0.a_{N-n}1)}\ket{1}\Bigr)
\otimes  
{1\over\sqrt2}\Bigl(\ket{0}+e^{2\pi i(0.a_{N-n-1}a_{N-n}1)}\ket{1}\Bigr)
\nonumber \\    
&\mbox{}&\hspace{-25pt}\hphantom{\ket{a_{N-n}\ldots}}
\otimes \cdots \otimes
{1\over\sqrt2}\Bigl(\ket{0}+e^{2\pi i(0.a_1\ldots a_{N-n}1)}\ket{1}\Bigr)
\nonumber \\
&\mbox{}&\hspace{-25pt}\hphantom{\ket{a_{N-n}\ldots}}
\otimes
e^{i\pi(0.x_1a_1\ldots a_{N-n}1)}
{1\over\sqrt2}\Bigl(\ket{0}+e^{2\pi i(0.x_1a_1\ldots a_{N-n}1)}\ket{1}\Bigr)
\;.
\end{eqnarray}
Since the quantum baker's map $\hat B_n$ maps the state (\ref{eqprod1}) to 
the state (\ref{eqprod2}), it can be seen that it shifts the states of all 
the qubits to the left, except the state of the leftmost, most significant 
qubit.  The state $|x_1\rangle$ of the leftmost qubit can be thought as being 
shifted to the rightmost qubit, where it suffers controlled phase changes
that are determined by the state parameters $a_1\ldots a_N$ for the original 
``momentum qubits.''  The quantum baker's map can thus be written as a shift 
map on a finite string of qubits, followed by controlled phase changes on the 
least significant qubit.

An important special case arises for $n=N$, for then there are no momentum
qubits on which to condition the phase changes of the least significant qubit.  
Working either from Eq.~(\ref{eqpartial}) or from Eqs.~(\ref{eqprod1}) and 
(\ref{eqprod2}), one can show that
\begin{equation}
\hat B_N\ket{x_1}\otimes\cdots\otimes\ket{x_N}=
\ket{x_2}\otimes\cdots\otimes\ket{x_N}\otimes
{e^{i\pi x_1}\over\sqrt2}\Bigl(
e^{-i\pi/4}e^{-i\pi x_1/2}\ket{0}+
e^{i\pi/4}e^{i\pi x_1/2}\ket{1}\Bigr) \;.
\end{equation}
The state $|x_1\rangle$ of the leftmost qubit is shifted to the rightmost 
qubit, where it undergoes a single-qubit transformation, not controlled 
by the state parameters of the other qubits.  As a result, this incarnation 
of the quantum baker's map, unlike the others, does not entangle initial 
product states.

In conclusion, we have given a symbolic representation of the states
of $N$ qubits that leads naturally to a class of
quantum baker's maps, defined as shift maps with respect to the symbolic
representation. For each of the maps in this class, there is a product
basis such that the action of the map on an arbitrary basis state is
equivalent to a shift of the string of qubits to the left, followed by
controlled phase changes on the rightmost qubit. This result is a
potential starting point for a generalization of the method of
classical symbolic dynamics to chaotic quantum maps.


\end{document}